\documentclass[a4paper, amsfonts, amssymb, amsmath, reprint, showkeys, nofootinbib, twoside, floatfix]{revtex4-1}

\usepackage[english]{babel}

\usepackage[utf8]{inputenc}
\usepackage{amsthm}
\usepackage{mathtools}
\usepackage{xcolor}
\usepackage{graphicx}
\usepackage[left=23mm,right=13mm,top=35mm,columnsep=15pt]{geometry} 
\usepackage{adjustbox}
\usepackage{placeins}
\usepackage[T1]{fontenc}
\usepackage{lipsum}
\usepackage{csquotes}
\usepackage{subfig}
\usepackage{comment}
\usepackage{float}
\usepackage{soul}

\usepackage[pdftex, pdftitle={Article}, pdfauthor={Author}]{hyperref}
\newcommand{\eps}{\varepsilon}
\newcommand{\dd}[1]{\text{d}#1}
\newcommand{\del}{\partial}
\newcommand{\dv}[2]{\frac{\dd{#1}}{\dd{#2}}}
\newcommand{\pdv}[2]{\frac{\del #1}{\del #2}}

\begin{document}
%\title{Numerical Investigation of the Logarithmic Schr\"{o}dinger Equation as a Wavefunction-Level Model for Decoherence and Localisation of a Quantum Particle}
\title{Numerical investigation of the logarithmic Schr\"{o}dinger model of quantum decoherence}

\author{Rory van Geleuken}
\author{Andrew V. Martin}
    \email[Correspondence email address: ]{andrew.martin@rmit.edu.au}
    \affiliation{School of Science, RMIT University, Melbourne, Victoria 3000, Australia.}

\date{\today}

\begin{abstract}
    A logarithmic Schr\"{o}dinger equation with time-dependent coupling to the non-linearity is presented as a model of collisional decoherence of the wavefunction of a quantum particle in position-space. The particular mathematical form of the logarithmic Schr\"{o}dinger equation has been shown to follow from conditional wave theory, but the validity of the logarithmic Schr\"{o}dinger equation has not yet been investigated numerically for general initial conditions. Using an operator-splitting approach, we solve the non-linear equation of motion for the wavefunction numerically and comparit it to the solution of the standard Joos-Zeh master equation for the density matrix. We find good agreement for the time-dependent behaviour of the ensemble widths between the two approaches, but note curious `zero-pinning' behaviour of the logarithmic Schr\"{o}dinger equation, whereby the zeros of the wavefunction are not erased by continued propagation. By examining the derivation of the logarithmic Schr\"{o}dinger equation from conditional wave theory, we indicate possible avenues of resolution to this zero-pinning problem.
\end{abstract}


\maketitle

\section{Introduction}
Understanding the behaviour of a quantum system under the continuous influence of its environment is of prime importance to a wide variety of research areas, ranging from fundamental questions about the quantum-to-classical transition \cite{Schlosshauer2004} \cite{Schlosshauer2007} \cite{Joos1985}, to the creation and operation of emerging quantum technologies \cite{Duan2001}. In particular, modelling the effects of decoherence by the environment is of vital importance in the design of these new technologies, where the loss of coherence between states is often a limiting factor \cite{Tyryshkin2012}. However, the macroscopic number of degrees of freedom of any physically reasonable environmental model results in analytically intractable models for the interactions, and so these must be treated approximately. Typically, a master equation is used, where the effects of a large class of environmental models, such as those obeying the Born and Markov approximations, can be incorporated on general principles by the addition of terms of Lindblad form \cite{Lindblad1976}. 

If we consider a quantum particle interacting with an environment of scattering particles, the Gallis-Fleming master equation \cite{Gallis1990} can be used to describe the environment's decohering effects in position-space. It is the loss of coherence between position space basis states that is responsible for the localisation of quantum particles, a feature that is completely absent in the bare Schr\"{o}dinger equation. The Gallis-Fleming master equation provides a very general description of this process and was re-derived and corrected by Hornberger and Sipe \cite{Hornberger2003}, after first being considered in approximate form by Joos and Zeh \cite{Joos1985}. This has been used to model the effects of environmental engineering in matter-wave interferometry \cite{Arndt2005}, to study the quantum-to-classical transition \cite{Schlosshauer2004} \cite{Zurek2003}, and in modelling the behaviour of quantum states of matter, such as Bose-Einstein Condensates \cite{Colombe2007}. 

Treating the environment as a bath of harmonic oscillators leads to the Quantum Brownian Motion (QBM) model, originally derived by Caldeira and Leggett \cite{calderia1983} using path integral techniques introduced by Feynman and Vernon \cite{vernon1963}, and solved analytically by Fleming et al \cite{fleming2011}. This model, specialised appropriately to the physical situation, has been used to study a variety of emerging quantum technologies, such as the decoherence of Josephson junctions \cite{Makhlin2001}, QED cavities \cite{Pellizzari1995}, and photonic devices \cite{Obrien2009}, among others \cite{Tian2002}.

A significant practical consideration in the implementation of these models is the computational complexity entailed by working with density matrices. The resources required to propagate such equations scale at least with the square of the size of the state space under consideration.

% Start Rev Response Addition
%============================

In a similar vein, the phase-space formalism, initially developed by Groenewald \cite{Groenewold1946} and independently by Moyal \cite{Moyal1949}, building on ideas from Wigner \cite{Wigner1932}, among others \cite{Curtright2012} involves converting the master equation formalism into a description of the evolution of quasiprobability distribution on phase space. This has been used very successfully to describe a wide variety of applications, such as the effects of decoherence in matter-wave interferometry \cite{Bateman2014}, modelling the behaviour of quantum walks as a basis for novel quantum algorithms \cite{Lopez2003}, modelling fundamental tests of quantum mechanics in cavity QED \cite{Milman05}, and even furnishing descriptions of coherent dynamics in the early universe \cite{Matacz94}. The flexibility of phase-space approaches is hindered only, as in the case of the density matrix formalism, by the growth of the computational complexity of numerical propagation in higher dimensional settings. This is because the Wigner-Weyl transformation between the density matrix description and the Wigner function description is invertible, and so both necessarily contain the same number of degrees of freedom.

The use of stochastic methods as a way of circumventing this computational scaling was pioneered by the quantum state diffusion methods of Gisin and Percival \cite{Gisin92}. They showed that the solution to a large class of master equations could be modelled by the evolution of a stochastic Schr{\"o}dinger equation with appropriately chosen stochastic forcing terms. Such calculations would result in a wavefunction that evolved in a non-deterministic manner, in response to the constant and random influence of the system's environment. The total history of a wavefunction calculated in such a way is known as a stochastic unravelling. In essence, these amount to samples in a Monte-Carlo numerical integration scheme of the path integral formulation, with each path corresponding to an unravelling, and can offer significant computational advantages over master equation methods \cite{Gisin93}. Additionally, conceptual insight into the interpretation of quantum mechanics with open systems offered by stochastic methods has been leveraged in proposals of modified theories of quantum mechanics such as the spontaneous collapse model of Ghirardi, Rimini and Weber \cite{Ghirardi86} and the subsequent continuous spontaneous localisation model of Ghirardi, Pearle, and Rimini \cite{Ghirardi90}. Extensions to relate these models to gravitation have also been proposed by Diosi \cite{Diosi87} and Penrose \cite{Penrose96}.

Even with these conceptual and numerical advantages in many applications, stochastic methods cannot guarantee an increase in performance in general. This is because for complex systems, the benefit of only having to propagate wavefunctions rather than density matrices is outweighed by the cost of requiring large sample sizes to ensure `weak' convergence (which is to say, convergence at the ensemble level, rather than convergence at the level of a single unravelling, or so-called `strong' convergence) \cite{Platen99}. Although various powerful schemes for reducing necessary sample sizes exist in specific applications of stochastic evolutions \cite{Shapiro2003}\cite{Kroese11}, there is no known general method for reducing the inherent statistical error below that guaranteed by the central limit theorem. For many applications of stochastic quantum propagation, this is not a problem, as the value in such methods often lies in (but is not limited to) the reproduction of realistic evolutions of a system of interest, for example. However, for the description of effects such as decoherence, a well-converged density matrix or equivalent is necessary, and in this respect, stochastic methods may not be able to out-perform the master equation formalism in general.

%============================
% End Rev Response Addition

Conditional wave theory (CWT) addresses this by working in a deterministic fashion\includecomment{in a deterministic fashion} at the wavefunction level, introducing the notion of marginal and conditional wavefunctions, which were first introduced by Hunter \cite{Hunter1975} and applied in the exact factorisation approach to molecular physics \cite{Abedi2012}. By exploiting the favourable computational scaling offered by this approach, as no sample averaging is necessary,\includecomment{as no sample averaging is necessary,} it may be possible to produce more computationally efficient models of decoherence.

% Start Rev Response Addition
%============================
The utility of this speed up is most evident when considering higher-dimensional systems. As noted above, when the physical situation can be reduced to a lower-dimensional problem, there exist powerful and efficient methods that CWT does not necessarily out-perform. However, as the number of spatial or abstract dimensions increases, working directly at the wavefunction level can significantly reduce computational overhead.

%============================
% End Rev Response Addition

This has noteworthy practical consequences for the simulation of matter-wave interferometry, for example, as working with the density matrix leads to a significant increase in the computational burden with increasing dimension. To take a concrete case, simulating a three-dimensional system using the master equation approach would require solution of a six-dimensional partial differential equation, whereas CWT reduces this to three dimensions.Other applications of our approach may exist in related fields of quantum matter, such as modelling the motion of solitons in BECs with dissipation or drag, or of the motion of quantum (quasi-)particles in other exotic media, which can prove challenging with increasing dimensionality.\includecomment{Other applications of our approach may exist in related fields of quantum matter, such as modelling the motion of solitons in BECs with dissipation or drag, or of the motion of quantum (quasi-)particles in other exotic media, which can prove challenging with increasing dimensionality.} 

The price of the reduction in dimensionality offered by CWT\includecomment{the reduction in dimensionality offered by CWT} is non-linearity of the resulting equations of motion. Nevertheless, this would still render previously impractical calculations tractable, allowing for experimental investigations of decoherence to be compared with theory.

It has been noted that both QBM and the scattering model yield the same dynamics in the limit of an environment dominated by long-wavelength (low momentum) and small central particle displacements. Their respective master equations reduce to the form originally derived by Joos and Zeh, which is itself in Lindblad form \cite{Schlosshauer2007}. This provides the motivation to construct the conditional wave theory corresponding to this important model. Consequently, a nonlinear equation of motion for the marginal wavefunction has been derived \cite{vang2020}, possessing a logarithmic non-linearity\includecomment{possessing a logarithmic non-linearity} which predicts the same evolution for as the Joos-Zeh master equation (JZME) for Gaussian states.

% Start Rev Response Addition
%============================
The Schr{\"o}dinger equation for a wavefunction $\psi$ augmented with a logarithmic non-linear term proportional to $-\psi\ln|\psi|$, known as the Logarithmic Schr{\"o}dinger Equation (LogSE) has been investigated as a possible model for a variety of non-linear phenomena. An early application was found in the description of the quantum-to-classical transition due to the separability property and existence of solitonic `Gausson' solutions \cite{Bialynicki76}\cite{Bialynicki1979}. Although this was not born out by experiments in atomic physics \cite{Shull80}, it found applications in nuclear physics \cite{Hefter1985} and in modelling solitonic behaviour of optically non-linear media \cite{Shen05}. It has also been proposed as a model of quantum information exchange \cite{Zlosh11} and Bose-Einstein condensation \cite{Avdeenkov11}, due to its obvious formal similarity to the expression for entropy. In contrast to these proposals, the logarithmic non-linearity arises from the CWT description of collisional decoherence, rather than as a speculative model.
%============================
% End Rev Response Addition

The aim of the current work is to investigate the LogSE as a model of decoherence for non-Gaussian wave-packets, using numerical techniques. In Secs. \ref{sec:theoryA} and \ref{sec:TimeDepNonLin} we briefly review the Joos-Zeh master equation and conditional wave theory, as well as the mathematical form of the corresponding LogSE predicted by the latter. In Sec. \ref{sec:reg} the regularisation of the logarithmic non-linearity is discussed and in Sec. \ref{sec:err}, important properties of error propagation in PDEs with time-dependent non-linear terms, such as the LogSE given by CWT, are discussed. Section \ref{sec:num} contains the results of our numerical investigation, with Gaussian (Sec \ref{sec:numgauss}) and non-Gaussian, double-peak (Sec \ref{sec:numlynch}) initial conditions presented. Section \ref{sec:numwidtherr} contains an analysis of the behaviour of the numerical error and the nature of the agreement between the JZME and LogSE. Section \ref{sec:disc} contains a discussion of the advantages and disadvantages of the methods presented, as well as in depth analysis of the curious `zero-pinning' behaviour exhibited by the LogSE in Sec \ref{sec:zeropin}.

\section{Theory}
\label{sec:theory}
\subsection{The Joos-Zeh Master Equation and Conditional Wave Theory}
\label{sec:theoryA}
A seminal result in the theory of quantum decoherence and localisation was the JZME. Joos and Zeh considered a massive particle with position $x$ that scatters massless (or approximately massless) environmental particles and found that the net effect of a macroscopic number of these events could be approximated by a simple quadratic decohering term. The regime in which the JZME's approximations hold is that of low-momentum environmental particles and small displacements of the central particle. With these conditions satisfied, the JZME governs the behaviour of the reduced density matrix $\rho_S(x,x')$ of a particle moving along the $x$-axis,
\begin{widetext}
\begin{equation}
    \frac{\dd{}}{\dd{t}} \rho_S(x,x',t) = \frac{i\hbar}{2m} \left(\frac{\del^2}{\del x^2} - \frac{\del^2}{\del x'^2}\right)\rho_S(x,x',t) - \frac{\Lambda}{\hbar}(x-x')^2\rho_S(x,x',t)
\end{equation}
\end{widetext}
where $\Lambda$ is the decoherence parameter, which controls the strength of the decohering effect of the environment. The reduced density matrix $\rho_S(x,x')$ is related to the total system-environment density matrix $\hat{\rho}$ through the partial trace,
\begin{equation}
    \rho_S(x,x') = \langle x ' |\text{tr}_{E} \hat{\rho}| x \rangle = \sum_{|e\rangle \in \mathcal{B}_E} \langle x ' | \otimes  \langle e | \hat{\rho} | e \rangle \otimes  | x \rangle,
\end{equation}
where $\mathcal{B}_E$ is an orthonormal basis of $\mathcal{H}_E$, the Hilbert space of environmental states, and $\otimes$ denotes the tensor product. Using the tools of conditional wave theory we have previously shown in a prior work \cite{vang2020} that the master equation formalism can be captured by a pair of coupled equations of motion for the conditional wavefunction, $\phi$, and marginal wavefunction, $a$. These are defined by a factorisation of the total system-plus-environment wavefunction $\psi$, through
\begin{equation}
    \psi(x,q,t) = \phi(x,q,t)a(x,t),
\end{equation}
where $x$ is the coordinate of the central particle of interest (the `system') and $q$ is an abstract coordinate that represents the configuration of the environment. The factorisation is not unique, and has an associated gauge symmetry realised by multiplying one factor by an $x$-dependent phase factor and the other factor by the conjugate of the same. Nonetheless, by choosing an appropriate gauge and constructing the marginal density matrix -- given by $\rho_m(x,x',t)=a^*(x',t)a(x,t)$ -- the resulting equation of motion can be compared to the JZME. We find that a Logarithmic Schr\"{o}dinger Equation of the form
\begin{equation} \label{eq:lse}
    i\dv{a(x,t)}{t} = -\frac{\hbar}{2m}\nabla^2a(x,t)+\frac{\hbar\gamma(t)}{m}a(x,t)\ln{|a(x,t)|^2}
\end{equation}
where $\gamma(t)$ is a real-valued function of time that reproduces the behaviour of solutions to the JZME in coordinate-space, provided that the environmental state is Gaussian in its coordinate. The details of the derivation can be found in the appendix.

The time-dependence of the coupling parameter, $\gamma(t)$ can be derived exactly for Gaussian initial conditions, and approximately generalised to non-Gaussian initial conditions, where it possesses asymptotically linear forms in both the long- and short-time regimes. As our aim is to solve Eq. \eqref{eq:lse} for non-Gaussian initial conditions, the asymptotic behaviour exhibited by $\gamma(t)$ results in a straightforward and computationally efficient numerical implementation. However, due to the nature of error propagation in non-linear PDEs, this linearly increasing coupling leads to numerical breakdown in finite time, as discussed in section \ref{sec:err}.

\subsection{Time Dependence of the Coupling to the Non-Linearity}
\label{sec:TimeDepNonLin}
As shown in the paper by Joos and Zeh \cite{Joos1985} where it first appeared, the JZME not only preserves the Gaussian functional form of states (so that initially Gaussian states remain Gaussian for all time) but also causes $O(t^3)$ spreading of the ensemble width, $w(t)$, of these states (which corresponds to the standard deviation of the associated probability distributions). For a free particle, the Schr\"{o}dinger equation predicts a linear growth rate for $w(t)$. Note that $w(t)$ is often referred to as the ensemble width, as it represents the standard deviation of a probability distribution and to distinguish it from the coherence length, which represents the distance over which a given state retains quantum phase information and hence can support superpositions. 

These two length scales can be more directly interpreted in terms of the density matrix. The coherence length measures the width of the distribution of off-diagonal entries, while the ensemble width measures the spread of the on-diagonal entries. For a free particle, these have identical time-dependence. Decoherence causes a reduction in the coherence length over time which is accompanied by an increase in the ensemble width. In reality, the coherence length will only fall to some minimum value related to the energy and length scale of the interaction between the system and environment. However, this minimum is closely related to thermalization, and will not appear in simplified models such as the JZME.

Substituting a Gaussian initial state into Eq. (\ref{eq:lse}), we find that the coupling parameter $\gamma(t)$ can be written as
\begin{equation} \label{eq:gammaintegral}
    \gamma(t) = \frac{2 \Lambda}{\hbar} \frac{1}{w(t)}\int_0^t \dd{t'}w(t').
\end{equation}
Working through the equations of motion, we find that $w(t) = O(t^3)$ and indeed the form of the resulting coupled differential equations for the parameters describing the Gaussians exactly agree \cite{vang2020}. Setting $w(0)=b$, the characteristic decoherence time is
\begin{equation}
    t_b = \frac{\hbar}{\Lambda b^2},
\end{equation}
which is the time taken for the coherence length to fall by a factor of $1/e$ (unless otherwise stated, we will be working in units where $b=\hbar=\Lambda=1$ for the remainder of this paper). The behaviour of $\gamma(t)$ can be split into two regimes, short- and long-times, like so,
\begin{align}
    \gamma(t) &= 2\Lambda t  + O(t^2), \qquad\qquad t <<t_b, \\
	\gamma(t) &= c_0 + \frac{\Lambda t}{2} + O(t^{-1}), \,\,\, \quad t >> t_b, 
\end{align}
where $c_0$ is a constant determined by the initial conditions. The transition between these two regimes occurs around the first decoherence time. In the numerical calculations, we used an interpolation between these two regimes rather than calculating the integral explicitly, even in the Gaussian case. The short time expansion clearly holds (by Taylor expanding the integral in Eq. \eqref{eq:gammaintegral} around zero, for instance) for an arbitrary time-dependent width (i.e. the second moment) of a general state, where that is well-defined. Although the long-time dependence is only exact in the case of Gaussian initial conditions, numerical investigation shows that it shows good agreement with the JZME in general. Furthermore, if we assume that the long time behaviour of the width is dominated by polynomial growth, so that $w(t)=O(t^n)$ for $t>>t_b$ then it's easily verified that
\begin{equation}
    \gamma(t)=\frac{2\Lambda}{\hbar} \frac{\int_0^t \dd{t'} w(t')}{w(t)} \propto \frac{t^{n+1}}{t^n} =O(t),
\end{equation}
which further motivates the general use of a linear time dependence in the limiting (long and short) regimes for a general state.
\begin{figure}
    \centering
    \includegraphics[width=0.45\textwidth]{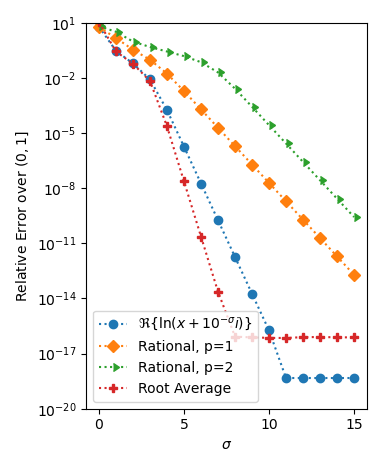}
    \caption{Relative $L^2(0,1]$ distance between the natural logarithm and a variety of regularised logarithms as discussed in the text. `Rational' refers to the form given in Eq. \ref{ratlog}, and `root average' refers to the form given in Eq. \ref{rootavg}.}
    \label{fig:regerr}
\end{figure}
\subsection{Regularisation of the Logarithm}
\label{sec:reg}
Although the limit,
\begin{equation}
    \lim_{z\rightarrow0} z \log |z| = 0
\end{equation}
for $z\in \mathbb{C}$, is straightforward to prove, it is necessary to regularise the logarithm for numerical stability. In particular we used the regularised log function, $\ln_\sigma$, defined as,
\begin{equation}
    \ln_\sigma(x) = \Re \ln(x+10^{-\sigma}i)
\end{equation}
for some $\sigma\geq0$ so that for $x>>0$,
\begin{equation}
    \ln_\sigma(x) \approx \ln(x)
\end{equation}
but as $x\rightarrow 0$, $\ln_\sigma(x)\rightarrow-\sigma\ln(10)$ rather than diverging to $-\infty$. In the results presented in this paper, $\sigma = 16$, as this was well below the level of desired numerical accuracy and did not noticeably effect the computational resources required. Other choices had little to no appreciable effect on the results, provided $\sigma \gtrsim 2$. Several other regularisation schemes were also trialled, including
\begin{equation} \label{rootavg}
    \ln_{(\sigma,N)}(x) = \frac{1}{N}\sum_{k=0}^{N-1} \ln\left(x+10^{-\sigma}e^{\frac{2\pi i k}{N}}\right),
\end{equation}
which also showed no appreciable sensitivity to choices of $\sigma$ or $N$ provided $\sigma\gtrsim3$. A slightly different approach took the form,
\begin{equation} \label{ratlog}
    \text{lnr}_{(\sigma,r)}(x) = \frac{x^p}{x^p+10^{-\sigma}}\ln(x+10^{-\sigma})
\end{equation}
which also did not display any appreciable effect on the evolution for $\sigma \gtrsim 3$ and all $p\geq 1$. Similar conclusions can be drawn from the graph in Fig. \ref{fig:regerr}, which shows the relative error in terms of the functional distance between the natural logarithm and the various regularisation schemes discussed. The functional distance was calculated using the formula
\begin{equation}
    \text{Err}(f,g) = \frac{||f-g||_2}{\sqrt{||f||_2 ||g||_2}}
\end{equation}
for two functions $f$ and $g$, where $||f||_2$ denotes the $L^2(0,1]$ norm, given by,
\begin{equation}
    ||f||_2 = \left( \int_0^1 \dd{x} |f(x)|^2 \right)^{1/2}.
\end{equation}

The fact that the singularity of $z \ln |z|$ at the origin is removable appears to be of much greater importance than the particular method of calculating the logarithm during the intermediate steps. This is consistent with the results of Bao et. al. \cite{Bao2019A}, which also showed that the evolution to the standard LogSE displayed insensitivity to the regularisation scheme used.

As discussed in more detail below, particularly in section (\ref{sec:zeropin}), the behaviour of solutions to the LogSE near zeros of the wavefunction disagree significantly with standard theory. However, the results in this section strongly suggest that these issues do not appear to be caused by the numerical details of the logarithmic regularisation, but analytic properties of the form of the LogSE itself and its derivation from CWT.

We used a split-operator approach with the second order Strang-splitting scheme to propagate the LogSE (see Bao et al \cite{Bao2019A} for a thorough analysis of operator splitting applied to the standard LogSE). This has been shown to have better performance both in computational time and reduced numerical error when compared to a (modified) Crank-Nicholson finite difference scheme which is standard for linear PDEs \cite{Bao2019B}. Operator splitting also has the advantage of being relatively straightforward to implement, as most of the computation is handled by Fourier transforms. In our case, the algorithm was implemented in Python, using the Numpy library \cite{numpy} for array manipulation and Fourier transforms. 

\subsection{Error Analysis} \label{sec:err}
%-Here we will discuss general numerical properties of non-linear PDEs and the implications for numerical stability.\\
%-This is covered in section 6.1 in the notes.

Although the Strang-splitting method has relatively low numerical error and is stable for the standard LogSE \cite{Bao2019A}, CWT requires the non-linearity to be time-dependent. This leads to an intrinsic source of numerical error for any discrete numerical scheme. To see this, consider a general non-linear Schr\"{o}dinger like equation,
\begin{equation}
     i\partial_t{u(x,t)} + \frac{1}{2} \nabla^2 u(x,t) = F[u(x,t)]
\end{equation}
where $F[u]$ is some arbitrary differentiable function with the square brackets denoting that $F$ will typically take a function as an argument, although it can also depend on $x$ and $t$ independently. Recall also that in our units $\hbar=m=1$. If we denote some approximate solution to this equation, given by some numerical scheme for example, by $\tilde{u}$ so that
\begin{equation}
    \tilde{u}(x,t) = u(x,t) + \eta(x,t),
\end{equation}
where $u(x,t)$ is the exact solution and $\eta(x,t)$ is the error.  If we try to evolve $\tilde{u}$ using the equation of motion, we find,
\begin{align} \label{eq:nlineom}
    i\partial_t u + \frac{1}{2} \nabla^2 u + i \partial_t \eta + \frac{1}{2} \nabla^2 \eta = F[u+\eta].
\end{align}
Expanding out the right hand side expression as a Taylor series in $\eta$,
\begin{equation}
    F[u+\eta] = F[u(x,t)] + \sum_{k=1}^{\infty} \frac{\del^k F[u]}{\del u^k} \frac{\eta(x,t)^k}{k!},
\end{equation}
and substituting this back into the right hand side of Eq. (\ref{eq:nlineom}), we can cancel out the exact solution's terms so we are left with,
\begin{equation} \label{eqn:errprop}
    i \partial_t \eta + \frac{1}{2} \nabla^2 \eta = \sum_{k=1}^{\infty} \frac{\del^k F[u]}{\del u^k} \frac{\eta(x,t)^k}{k!},
\end{equation}
which means that the error of any approximate solution to a non-linear equation propagates according to a non-linear equation of its own. For equations with a non-linearity that is polynomial in the intensity, such as the Gross-Pitaevskii Equation (GPE), the right hand side of (\ref{eqn:errprop}) will be of finite order. However, in general the sum will extend over infinitely many terms. Nonetheless, if $\tilde{u}$ is a good approximation, then $\eta$ can be taken to be small everywhere and we can truncate the series at $k=1$. That is,
\begin{equation} \label{eq:firstborn}
    \left(i\del_t + \frac{1}{2}\nabla^2\right)\eta(x,t) = F'[u] \eta(x,t) + O(\eta^2),
\end{equation}
where $F'[u]=\del F/\del u$. The differential operator on the left hand side of (\ref{eqn:errprop}) shares a Green's function with the homogeneous time-dependent Schr\"{o}dinger equation over $\mathbb{R}^d$,
\begin{equation}
\label{eq:greens}
    G(x,t) = i\Theta(t)\left(\frac{i}{2\pi t}\right)^{d/2} e^{-i \frac{x^2}{2t}},
\end{equation}
where $\Theta(t)$ is the Heaviside step function, such that,
\begin{equation} 
    \left(i\del_t + \frac{1}{2}\nabla^2\right)G(x,t) = \delta^d(x)\delta(t).
\end{equation}
Although equation (\ref{eq:firstborn}) cannot be solved exactly either, we can extract some information about the behaviour of the errors propagated by (\ref{eqn:errprop}) by expanding the solution in a series, $\eta(x,t)=\sum_n\eta_n(x,t)$, where,
\begin{equation} \label{eqn:errborn}
    \eta_{n+1} = G\ast(F'[u]\eta_n)
\end{equation}
and $\ast$ denotes convolution. The first term, $\eta_0$, can be thought of as the error intrinsic to the original approximation, and $\eta_n$ for $n\geq1$ represents error propagated due to the non-linear nature of the equation of motion. Since the Green's function contains a factor of $t^{-d/2}$, we expect that the higher order propagated errors will die out quickly. This is indeed the case for the GPE, were $F'[u]$ is not explicitly time dependent. However, in the case of CWT, we have a time-dependent coupling to the non-linearity, and so $F'[u]$ has an approximately linear time dependence. Thus, each iteration of (\ref{eqn:errborn}) yields a factor of $t^{1-d/2}$. For $d=1$, this means that the propagated errors will eventually diverge. For $d=3$, we would expect these errors to die out, whereas for $d=2$, a more subtle analysis is required. These latter two cases are outside the scope of this paper, though the methods we present can be generalised to higher dimensions and are intended to form the basis of future work.

While we have shown that any numerical scheme for evolving our equation of interest is inherently unstable, we found that the Strang-splitting approach was stable over the time-scale required.

\,

\begin{widetext}

\begin{figure}
    \centering
    \includegraphics[width=0.45\textwidth]{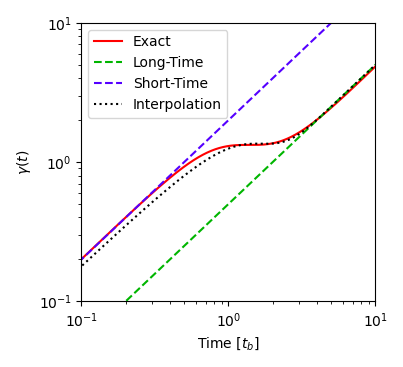}
    \caption{The behaviour of $\gamma(t)$ for a Gaussian wave packet over the timescale of interest, along with its interpolated linear approximation. Note that this is a log-log plot so straight lines here correspond to straight lines on the equivalent rectangular axes, but offset lines have different gradients, even if they appear parallel.}
    \label{fig:linapprox}
\end{figure}
\section{Numerical Results}
\label{sec:num}
\subsection{Gaussian Initial Conditions}
\label{sec:numgauss}

%-In this section we'll lay out the results for the Gaussian case along with the relative error of the linear interpolation for the coupling to the non-linear term. \\
%-We will also demonstrate the agreement between the JZME and the LogSE for the ensemble width for non-Gaussian wavefunctions.\\
%-This will link into the discussion in the second half or so of the paper about zero-pinning.
As demonstrated in \cite{vang2020}, CWT can exactly reproduce the evolution predicted by the JZME for Gaussian initial conditions. Moreover, CWT allows us to make a so-called `linear-time' approximation, which simplifies calculations at the cost of accuracy. Even with this simplified approach, we find that some qualitative features are captured by the LogSE. Recalling the discussion in section \ref{sec:TimeDepNonLin}, we note that since the coupling to the non-linearity transitions between two distinct linear regimes, we try an interpolation between them to approximate $\gamma(t)$. That is, our $\gamma(t)$ is given by
\begin{equation}
    \gamma(t) = 2\Lambda t (1-\sigma(t)) + \left(c_0 + \frac{\Lambda t}{2}\right) \sigma(t),
\end{equation}
where $\sigma(t)$ is the function
\begin{equation}
    \sigma(t) = \frac{1+\tanh(t-t_{b})}{2}.
\end{equation}
Any sigmoid function can be used for $\sigma(t)$ and the hyperbolic tangent was selected as it offered a good combination of both smoothness of transition and simplicity of implementation. Other widths of the transition (that is, scaling of the time coordinate) can be used, but since we are working in units natural to the problem, the simplest form provided the best agreement.

This linear approximation for $\gamma(t)$ has been plotted in Fig. \ref{fig:linapprox}. In this manner, we smoothly interpolate between the short- and long-time regimes with a transition over the characteristic decoherence time, $t_b$. Naturally, this does introduce some error, even in the Gaussian case, which can be seen as a slight disagreement between the exact JZME solution and the LogSE solution, which is visible in Fig. \ref{fig:gauss2x2}. On the other hand, this approach is reasonably general and simple, as well as being quite flexible, as different interpolation methods could produce better agreement in regions of interest (much shorter than the decoherence timescale or much longer, for example).

\begin{figure}[H]
    \centering
    \includegraphics[width=0.85\textwidth]{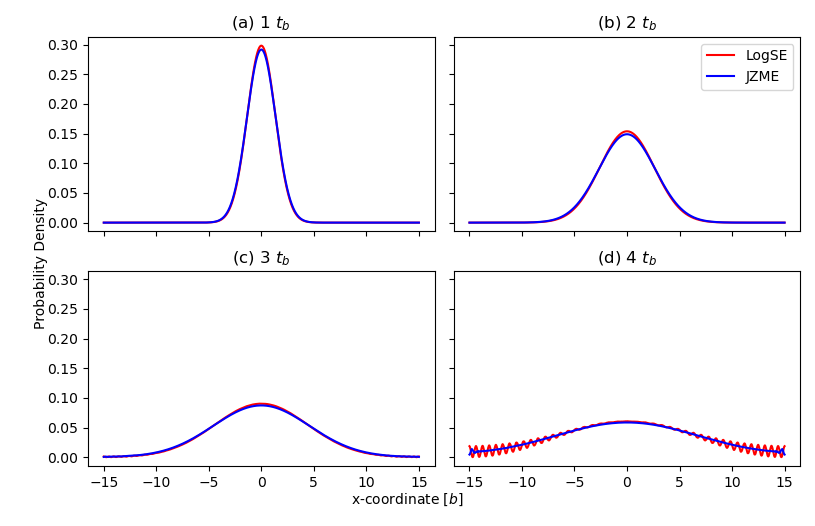}
    \caption{Gaussian initial conditions propagated both by the JZME and the LogSE with the linear-interpolation form $\gamma(t)$ function at multiples of the characteristic decoherence times. This demonstrates the good agreement between the two models, until the wave-packet encounters the boundary. The origin of the interference fringes is discussed in the main text.}
    \label{fig:gauss2x2}
\end{figure}

Figure \ref{fig:gauss2x2} was created using a time-step of 0.05 times the characteristic timescale (which in our units is 1), a domain of width 30 comprising 2048 points (or a spatial step size of approximately 0.015). The Gaussian initially had a unit standard deviation and a mean of zero.

The apparent interference effects evident after about 4 decoherence times in Fig. \ref{fig:gauss2x2} are due to the periodic boundary conditions implicit in the Fourier transforms used to perform the split-operator steps. This leads to one side of the Gaussian wrapping around the domain and interfering with the other half once it has grown sufficiently wide. This can be postponed by simply using a wider domain, although at the expense of a larger number of spatial steps to maintain the same density of points.

% Start Rev Response Addition
%============================

\subsection{Non-Gaussian, Zero-Free Initial Conditions}

Two non-Gaussian functions that do not possess zeros were also trialled as initial conditions. These were the Lorentzian, given by,
\begin{equation}
    a(x,0) = \frac{1}{\sqrt{\pi}} \frac{1}{1+\left(\frac{x}{b}\right)^2}
\end{equation}
for a `width' parameter $b$, and the hyperbolic secant, given by,
\begin{equation}
    a(x,0) = \frac{1}{b}\mathrm{sech}\left(\frac{x}{b}\right) = \frac{2}{\sqrt{b}} \frac{1}{\exp\left(\frac{x}{b}\right)+\exp\left(-\frac{x}{b}\right)}
\end{equation}
with analogous width parameter $b$. Note that the width parameter for the Lorentzian represents half the full width at half maximum (FWHM) and not the standard deviation, since the latter is ill-defined for a Lorentzian distribution.

\begin{figure}[p]
    \centering
    \includegraphics[width=0.8\textwidth]{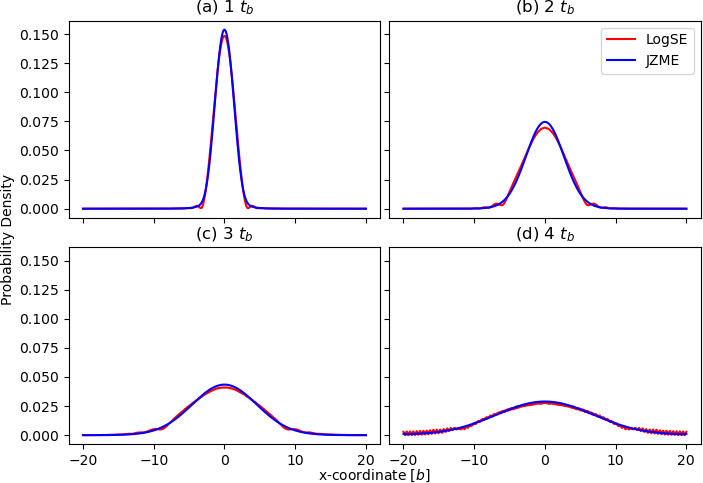}
    \caption{An initially Lorentzian wavefunction evolving under both the JZME and LogSE with the linearly-interpolated for of $\gamma(t)$ at multiples of the decoherence time. Although zero-pinning effects, discussed in detail in Sec. \ref{sec:zeropin}, are visible, the ensemble width shows good agreement between the two methods.}
    \label{fig:lrntz}
    \centering
    \includegraphics[width=0.8\textwidth]{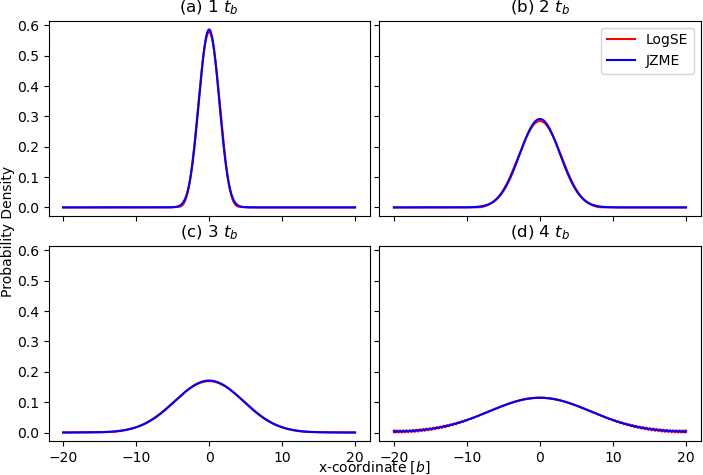}
    \caption{A wavefunction initially described by the hyperbolic secant evolving under both the JZME and LogSE with the linearly-interpolated for of $\gamma(t)$ at multiples of the decoherence time. Since zeros do not develop during evolution, the agreement between the two methods is very good until finite domain effects cause spurious interference effects, similar to the Gaussian case. The origin of these effects is discussed in the main text.}
    \label{fig:sech}
\end{figure}
\newpage

%============================
% End Rev Response Addition

\subsection{Double Peak Initial Conditions}
\label{sec:numlynch}
%-Here will show and discuss the behavior of the relative error for some other non-Gaussian initial conditions.\\
%-Notably, some initial conditions result in the LogSE solution oscillating around the JZME solution, which warrants further investigation.
For non-Gaussian initial conditions, we observe a phenomenon we have termed `zero-pinning' whereby points of zero intensity in the wavefunction are preserved by the evolution under the LogSE. This is particularly evident in Fig. \ref{fig:lynch2x2}, where the initial marginal wavefunction is a coherent superposition of two Gaussians, of the form, \includecomment{marginal wavefunction is a coherent superposition of two Gaussians, of the form,}
\begin{equation}
    a(x,0) = N(b,s)^{-1/2}\left[\exp\left(-\frac{(x-s)^2}{4b^2}\right)+\exp\left(-\frac{(x+s)^2}{4b^2}\right)\right]
\end{equation}
where $N(b,s)$ is a normalisation factor. We chose $s=b=1$ in our units, so the peaks are initially of unit width and separated by two units.\includecomment{where $N(b,s)$ is a normalisation factor. We chose $s=b=1$ in our units, so the peaks are initially of unit width and separated by two units.} Again, we used a time-step of 0.05 and a spatial step size of approximately 0.015 (a 30 unit wide domain with 2048 points).

As the two Gaussian peaks spread out, they begin to interfere, creating zeros in the probability distribution. The zeros spread out at the correct rate, but their visibility is not diminished, unlike solutions to the JZME. This represents a significant drawback to using the LogSE as a model of decoherence, as decreased fringe visibility is a central feature of decoherence, and is used extensively in experimental work to characterise the decohering effect of the environment \cite{Arndt2005}. The analytic nature of this problem is discussed further in section \ref{sec:zeropin}.

\begin{figure}[H]
    \centering
    \includegraphics[width=0.9\textwidth]{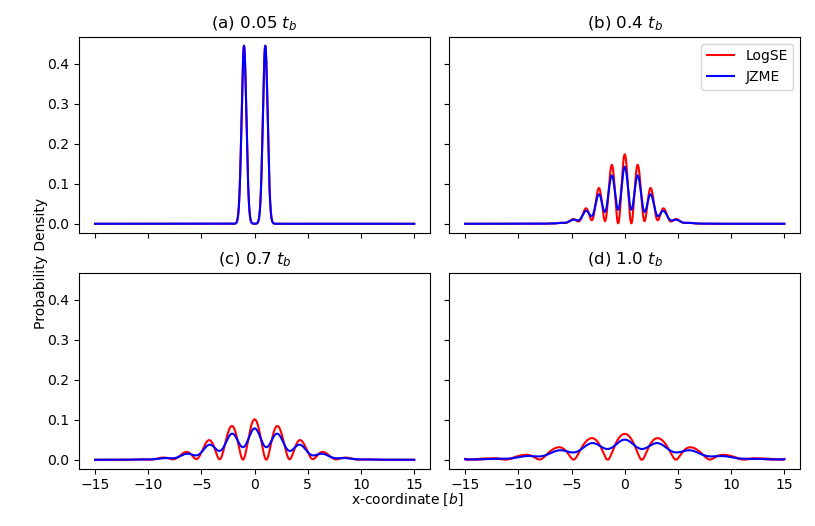}
    \caption{
    Twin Gaussian initial conditions at various fractions of the characteristic decoherence time. As in Fig. (\ref{fig:gauss2x2}), the LogSE is implemented with the linear-interpolation form of $\gamma(t)$. This demonstrates the good agreement in the ensemble width behaviour of the LogSE solution, while the `zero-pinning' effects, discussed further in Sec. \ref{sec:zeropin}, are evident.}
    \label{fig:lynch2x2}
\end{figure}

In all test cases where zeros were present, zero-pinning was observed. There is no sensitivity to the origin of the zero, whether it is a feature of the initial conditions, interference effects (such as the twin-Gaussian case shown above), or as a result of dynamics (for example, an isolated Cauchy-Lorentz distribution will spontaneously form zeros as it spreads out). As discussed below, this is likely a result of the analytic properties of the LogSE and not a numerical artifact. Moreover, even with the zero-pinning, all test cases showed good agreement with the predictions of the JZME regarding the time-dependence of the ensemble widths of the distributions. 

Note that since the same numerical integrator was used for both the JZME and LogSE, the JZME solutions simulated here also have implicitly periodic boundary conditions. Their interference effects are not visible here, however, since the JZME does not exhibit zero-pinning and appears to be somewhat more efficient at suppressing the off-diagonal terms in the density matrix. 

Conversely, the use of the same integrator for both formalisms resulted in a good demonstration of the significant computational speed-up offered by the LogSE model. Performing the calculations (on a commercial PC CPU) to produce the figures in this paper took on the order of hours for the highest spatial resolution calculations\includecomment{for the highest spatial resolution calculations} using the JZME but took only a few minutes using the LogSE. 

%Although operator splitting is not necessarily the most efficient method for JZME, since it is linear in the density matrix which means that finite-difference methods may be more efficient, this reduction in computation time demonstrates an important advantage of the LogSE model. 

However, operator splitting is not necessarily the most efficient method for propagating the JZME, as it is linear in the density matrix. Generally, finite-difference methods (such as the well-established implicit Crank-Nicholson method) may exhibit better performance for these types of partial differential equations. Nonetheless, the observed reduction in computation time still serves as a good comparison to demonstrate an important advantage of the LogSE model.
\end{widetext}

\subsection{Widths and Error Growth}
\label{sec:numwidtherr}

Although solutions to the LogSE display different behaviour to the corresponding JZME solutions, there is good agreement in the behaviour of the ensemble-width spreading, as shown in Fig. \ref{fig:widths}. Note that the `kink' present around $0.1t_b$ corresponds to the initial moment the interference fringes form, which effects the growth rate of the width. This is because the fringes contain less probability mass far away from the origin, and instead concentrate it in the central band.

As discussed in section \ref{sec:err}, the growth in numerical error for a 1-dimensional simulation of the LogSE grows over time regardless of numerical details. At the same time as this error is being propagated, the typical (absolute) value of the wavefunction is shrinking due to the rapid ensemble width growth (rapid compared to the free Schr\"{o}dinger equation solution with the same initial conditions, that is). Since the logarithm rapidly approaches an asymptote for real arguments in $(0,1]$, the function is quite steep in this region. Consequently, small changes in the absolute value of the wavefunction can have a considerable effect on the value of the logarithm, which may contribute to numerical instability.

Thirdly, the self-interference caused by the periodic boundary conditions can occur quite early in the simulation due to the rapid spreading out of the wavefunction. Combined with the zero-pinning, these effects can be difficult to tease apart by varying the parameters of the simulation (temporal and spatial resolutions, domain size, etc). 

Incorporation of absorptive boundary conditions was attempted in order to mitigate the effects of self-interference, but were ultimately disregarded. An imaginary potential that was non-zero only at the boundary was trialed, however, unless the width of this non-zero region was carefully tuned reflections from the boundary were produced (\cite{he2007} contains a good discussion of this problem). This tuning seemed to be sensitive to the initial conditions, unlike in the linear case, and simply introduced a new problem while not completely eliminating the wrapping of the wavefunction around the domain. While more mathematically rigorous and physically realistic methods exist for implementing absorptive boundary conditions, these can introduce significant computational complexity because they require an additional integral to be performed at each time-step, as well as being non-local in time (see \cite{Lubich2003} for an example algorithm and general analysis).

Ultimately, the computational scaling of integrating the LogSE with Strang-splitting means that it is, in general, much more tractable to simply work with a very large domain to mitigate the spurious effects of periodicity.

Regardless, the good agreement between the two formalisms (master equation versus LogSE) as far as the ensemble width is concerned suggests that while the form of the LogSE that has been derived captures some aspects of the environmental influence, the mechanism by which fringe visibility is diminished is lost. Interestingly this also suggests that these are - in some sense - separate phenomena, which is not obvious \textit{a priori}, since both appear simultaneously with the addition of the single decohering term in the JZME.
\,
\begin{widetext}

\begin{figure}
    \centering
    \includegraphics[width=\textwidth]{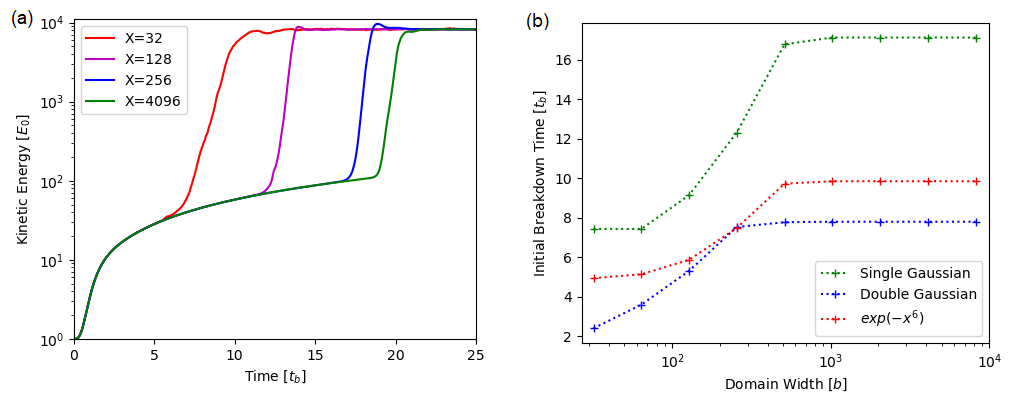}
    \caption{(a) The kinetic energy of a Gaussian state plotted against time, in units of the initial kinetic energy. Initially, the kinetic energy grows linearly (note the logarithmic axis) and suddenly jumps when the wavepacket encounters the edge and wraps around the domain. (b) The point at which this jump first occurs, plotted with increasing domain width. Note that for domains wider than about $10^3b$, the numerical breakdown is no longer due to self-interference, but is intrinsic to the LogSE, as implemented here.}
    \label{fig:DomsEnergy}
\end{figure}

\begin{figure}
    \centering
    \includegraphics[width=\textwidth]{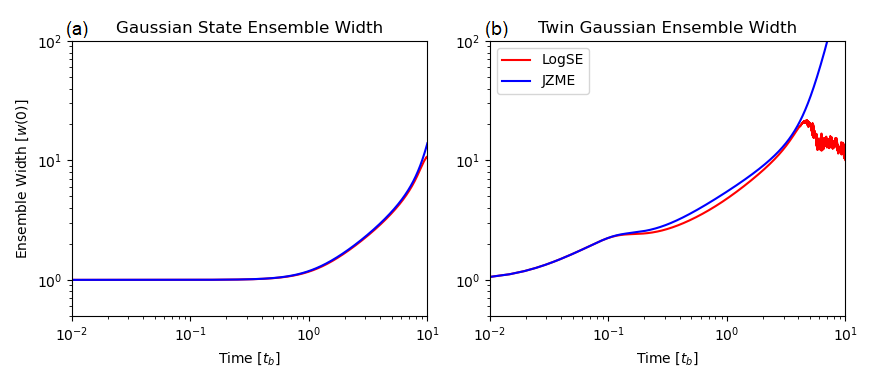} 
    \caption{Plots of the ensemble widths of the probability distributions corresponding to the Gaussian initial state (left) and the twin-Gaussian initial state (right), measured relative to the initial width of the distribution ($w(0)=b$). Note that the `kink' that occurs in the twin-Gaussian plot at around $10^{-1}t_{b}$ corresponds to the initial formation of the interference fringes. The divergence of the widths just before $10 t_{b}$ corresponds to the numerical breakdown discussed in section \ref{sec:err}.}
    \label{fig:widths}
\end{figure}

\begin{figure}[ht]
\includegraphics[width=0.7\textwidth]{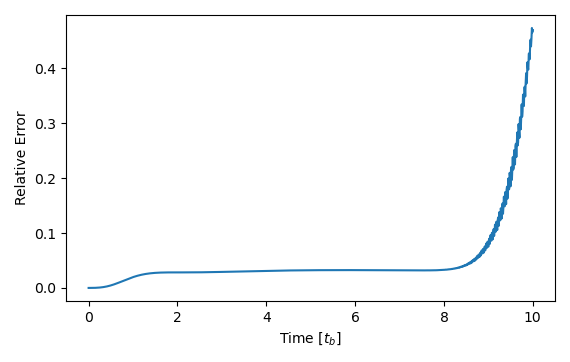}
\caption{The relative $L^2$ error (defined in the text), of the JZME and LogSE evolutions of a Gaussian initial state. The rapid increase in error around 9 characteristic times is insensitive to domain width and spatial resolution, corresponding to the inevitable numerical breakdown discussed in section \ref{sec:err}.}
\label{fig:errs}
\end{figure}

\section{Discussion}
\label{sec:disc}
As shown in Fig. \ref{fig:errs} (a), the LogSE performs very well as a model of decoherence for a Gaussian initial state. In all other test cases investigated, the LogSE yielded ensemble width growth that closely matched the JZME solution, before numerical instability set in. Since the LogSE's computational demand scales much more favourably than the master equation approach, it is not unreasonable to decrease the step size somewhat in order to improve stability. However, as shown in \textbf{Fig. \ref{fig:DomsEnergy}}, this only delays the breakdown to a point. As discussed in Sec. \ref{sec:err}, the time dependence of the non-linearity all but guarantees a finite-time numerical breakdown. 
%Interestingly, this also seems to produce stable states which spontaneously evolve out of the noisy regime that forms post-breakdown. These can be seen in figure [placeholder]. The stable forms appear to be roughly Gaussian in shape, and may have some relationship to the known solitonic solutions \cite{Bialynicki1979} to the LogSE. However, sensitivity to computational parameters like temporal and spatial resolution suggests that it's possible it may be an artifact of the operator splitting approach.

Furthermore, our analysis indicates that in more than one dimension, the accumulation of error intrinsic to the non-linearity of the problem should not prove as destructive. As discussed above, this is a consequence of the form of the Green's function in equation \ref{eq:greens}. This is of particular interest as the computational speedup offered by conditional wave theory is also most evident in higher dimensional problems. This suggests that the application of conditional wave theory to quantum mechanical problems involving continuous degrees of freedom, such as matter-wave interferometry or decoherence of quantum states of matter could be a promising area for further research.
Besides ensemble width growth, the other major prediction of decoherence theory is the loss of fringe visibility in interference patterns. As discussed in the results section, the LogSE in this form cannot reproduce this behaviour. This is a significant hurdle in the way of using this model in practical situations such as prediction matter-wave interference patterns. The cause of this problem is discussed in the next section, and possible areas of investigation which may offer a solution are proposed.

\subsection{Zero-pinning}
\label{sec:zeropin}
We now discuss in depth the mathematical origins of the zero-pinning phenomena as this is the most significant drawback of the LogSE model. Furthermore, the way that the LogSE treats zeros so differently to other values of the wave-function suggests that some important mathematical structure is missing or incorrect in the derivation of the LogSE from CWT. Although several approximations have been made, it's not immediately apparent how these would result in the observed effects. For example, most of the approximations make no reference to the on-diagonal values of the density matrix, or are concerned with simplifying the time dependence of the non-linear coupling. Although the zero-pinning is time-independent (in that it occurs regardless how far along in time the simulation is) we will show that the assumption that the evolution is dependent on purely local information contained in the wavefunction is closely related to the appearance of zero-pinning effects.

Assume that at some point on the $z$-axis, say $2x_0$, the intensity $\rho(x_0,x_0)$ goes to zero. We now consider why the addition of a term proportional to $y^2\rho(y,z)$ in the master equation should fill in the zero when this term clearly vanishes at this point. To begin, we note that along the $z$-axis $\rho$ must be real and a short distance along this axis - say $\delta x$ - from the point $x_0$, $\rho$ must be small, positive, and approximately parabolic. This follows if the zero is of first order in the wavefunction.
If we Taylor expand in time around the zero (assuming $\rho$ vanishes at some time $t$), we find,
\begin{equation} \label{eqn:rhoexpand}
    \rho(x_0,x_0,t+\dd{t}) = \dd{t}\left(\frac{i}{\mu}\del^2_{yz}\rho - \lambda y^2 \rho\right) + \frac{\dd{t}^2}{2} \left( \frac{i}{\mu}\del^2_{yz}\dot{\rho} - \lambda y^2 \dot{\rho} \right) + \frac{\dd{t}^3}{3!} \left( \frac{i}{\mu}\del^2_{yz}\ddot{\rho} - \lambda y^2 \ddot{\rho} \right) + O(\dd{t}^4),
\end{equation}
\end{widetext}
where $\mu^{-1} = 2\hbar/m$, $\lambda=\Lambda/\hbar$, the dots denote differentiation with respect to time, and everything on the right hand side is understood as being evaluated at the point $(x_0,x_0)$ at time $t$. We can find the expressions for the higher time derivatives of $\rho$ by repeated application of the JZME (Eq. (\ref{eqn:JZME})). A significant amount of algebraic work can be saved by noting that most of the terms generated in this manner will vanish. Trivially, anything with a surviving factor of $y$ must be zero, since this clearly vanishes at $(x_0,x_0)$. Furthermore, the derivative operator can be rewritten
\begin{equation}
    \del^2_{yz}\rho = \frac{1}{4} \left(\del^2_x\rho-\del^2_{x'}\rho\right)
\end{equation}
but since $\rho$ vanishes along both of the axes $(s,x_0)$ and $(x_0,s)$ (where $s\in\mathbb{R}$) the second derivatives with respect to both $x$ and $x'$ must both vanish at $(x_0,x_0)$. From all this, we can conclude that both the first- and second-order-in-time terms vanish, but the third order term will contain the following,
\begin{equation}
    \dddot{\rho} \sim \left(\frac{i}{\mu}\frac{\del^2}{\del y \del z}\right)^2\left(-\lambda y^2 \rho\right) \sim \frac{2\lambda}{\mu^2} \frac{\del^2\rho}{\del z^2} \approx \frac{16\hbar\Lambda}{m} \frac{\rho_\delta}{\delta x^2},
\end{equation}
where $\sim$ denotes equality up to cancellation of terms that vanish according to our observations above and $\rho_\delta$ is the value of $\rho(x,x)$ a short distance in either direction along the $z$-axis. As we noted, $\rho$ is approximately parabolic along this axis, and so the second derivative along it is positive (since $\rho$ must be non-negative along $z$). Thus for short times, the growth of absolute value of the wavefunction will be cubic in time.

Note this argument depends crucially on the fact that the JZME is manifestly non-local. Information about the state at the point $x'$ is propagated to $x$ through the decoherence term encountering the kinetic term. In the LSEs that we have developed so far, all information remains local, and so cannot be propagated to fill in the zeros.

A similar analysis can be performed for the LogSE. To begin, the marginal wavefunction can be expanded over some small time-step $\dd{t}$ at some first-order zero $x_0$ and eliminating terms that vanish at the zero,
\begin{equation}
    a(x_0,t+\dd{t}) = \dd{t}^2\frac{1}{m} \nabla \varepsilon \cdot \nabla a + O\left(\dd{t}^3\right),
\end{equation}
recalling that $\nabla^2a = 0$ at a first order zero. Furthermore, from the derivation of the LogSE (detailed in the appendix) the gradient of $\varepsilon(x)$ is given by,
\begin{equation}
    \pdv{\varepsilon(x)}{x} = -\frac{\hbar^2}{m} \frac{1}{|a(x)|}\frac{\del}{\del x}\{|a(x)|\gamma(x)\}
\end{equation}
so that,
\begin{equation}
    a(x_0,t+\dd{t}) = -\frac{\dd{t}^2}{2}\frac{\hbar^2}{m^2} \frac{1}{|a|}\frac{\del}{\del x}\{|a(x)|\gamma(x)\} \frac{\del a}{\del x}
\end{equation}
setting $\hbar=m=1$, the coefficient of $\dd{t}^2$, $c_2$, can be written,
\begin{equation} \label{eq:c2def}
    c_2 = \pdv{\gamma}{x}\pdv{a}{x} + \frac{\gamma}{|a|}\pdv{|a|}{x}\pdv{a}{x},
\end{equation}
which appears to be singular at the zero, since $|a|\rightarrow0$. However, care must be taken, since that is also precisely the point at which $|a|$ is not differentiable. Expanding over some small step $\dd{x}$ using a central finite difference and setting $|a(x)|=r(x)$,
\begin{align}
    &\lim_{x\rightarrow x_0} \frac{1}{|a(x)|} \pdv{|a(x)|}{x} \\ 
    &= \lim_{x\rightarrow x_0}\frac{1}{r(x)}\lim_{\dd{x}\rightarrow 0} \frac{r(x+\dd{x}/2)-r(x-\dd{x}/2)}{\dd{x}}\\
    &= \lim_{x\rightarrow x_0}\frac{r'(x_0^+) + r'(x_0^-)}{r(x)}.
\end{align}
Note that close to a zero, $f'(x_0^+)=-f'(x_0^-)$ for any continuous real function $f$. So it must hold that,
\begin{equation}
    \lim_{x\rightarrow x_0} \frac{1}{|a(x)|} \pdv{|a(x)|}{x} = \lim_{x\rightarrow x_0}\frac{0}{r(x)} = 0.
\end{equation}
The second term of $c_2$ in Eq. \eqref{eq:c2def} can therefore be dropped, leaving,
\begin{equation}
    a(x_0, t+\dd{t}) = -\dd{t}^2\frac{\hbar^2}{m^2} \gamma'(x_0, t) a'(x_0, t),
\end{equation}
which is clearly non-zero if $a$ is not everywhere zero and $\gamma$ is a function of space. In the simplest version of the CWT LogSE that we've used so far, $\gamma$ is not a function of space, implying that $\gamma'(x)=0$ which in turn means the zeros cannot be filled in.

Following the derivation of the LogSE given in the appendix, we can arrive at a slightly more general solution for the non-linear term than is given in Eq. (\ref{eq:lse}), 
\begin{equation}
\label{eq:eps1}
     \eps(x,t) = C(t) - \int_0^{x} \dd{x'} \frac{1}{|a(x',t)|} \frac{\del(\gamma(x',t)|a(x',t)|)}{\del x'},
\end{equation}
where $C(t)$ is an arbitrary function of time that does not affect the dynamics. If we allow $\gamma$ to be independent of $x'$ then we recover our expression for the logarithmic non-linearity. In general however, $\varepsilon$ can be written,
\begin{equation}
\label{eq:eps2}
    \varepsilon(x,t) = -\gamma(x,t) - \int_0^x \dd{x'} \gamma(x',t) \frac{\del}{\del x'} \ln |a(x',t)|
\end{equation}
which follows from the product rule applied to the integrand of Eq. (\ref{eq:eps1}) and the constant of integration has been set to zero. This form suggests that the influence of the coupling to the environment manifests in two ways. The first term represents a local potential which arises directly from the interaction with the environment and a second, non-local term which is difficult to interpret physically. Sergi and Zloschastiev \cite{Sergi2013} have discussed the logarithmic Schr\"{o}dinger equation in the context of quantum information transfer, and the above form of the non-linear term suggests a kind of weighted sum over the differential information content of the probability distribution generated by the marginal wavefunction. Thus the integral term may be interpreted as the amount of entanglement with the environment, as measured by the delocalised quantum information content of the marginal state.
The principal challenges in applying Eq. \eqref{eq:eps2} unsurprisingly lies in evaluating the integral. Firstly, at points where the magnitude of the wavefunction vanishes, the logarithm is ill-defined. However, this does not mean that the integral itself is undefined. If we regard $x$ as the real part of a complex coordinate, it is possible to evaluate the integral assuming the marginal magnitude is meromorphic on an appropriate domain. This can be done using the generalised Cauchy argument principle, which states that for a meromorphic function $f$ and a holomorphic function $g$ of a complex variable $w$, defined on an open subset of the complex plane $D$,
\begin{equation}
    \frac{1}{2\pi i} \oint_C \dd{w} \frac{f'(w)}{f(w)} g(w) = \sum_z g(z) n_C(z) - \sum_p g(p) n_C(p),
\end{equation}
where $C$ is a closed curve in $D$ that does not intersect the zeros ($z$) or poles ($p$) of $f$, and $n_C(x)$ is the winding number of the point $x$ with respect to the curve $C$ (note that if $x$ does not lie in the interior of $C$, $n_C(x)=0$).
Currently, we know of no sufficiently general solution for $\gamma(x,t)$ to actually apply this formula or how to find such a solution numerically. Nonetheless, the importance of zeros in the generalized Cauchy argument principle, along with the seemingly reasonable assumption that there should not be any poles in $|a(x,t)|$ on the real line is suggestive of a possible resolution to the zero-pinning problem.

\section{Conclusion}

Conditional wave theory offers new mathematical tools for the analysis of quantum decoherence, a central and important problem in the era of emerging quantum technologies. We have shown that as applied to the localising effects of decoherence, as measured by the behaviour of the ensemble width of a state, the logarithmic equation of motion predicted by CWT demonstrates excellent agreement with standard theory.
However, the mathematically simplest form of this equation of motion leads to significant accumulation of error and zero-pinning behaviour. We have shown that the former is accompanied by intrinsic reduction of necessary computational resources, by working at the wavefunction level and becomes less problematic with increasing dimension.
The zero-pinning effect prevents the destruction of interference effects that are well-known to occur as a result of decoherence. The resolution of this shortcoming of CWT may lie in careful analysis of the subtle non-local behaviour of the coupling to the environment, which appears to have mathematical links to previous work on information theory and the logarithmic Schr\"{o}dinger equation.

\appendix*

\section{Derivation of the LogSE} \label{sec:devapx}
The one-dimensional JZME for the reduced density matrix $\rho_S(x,x')$ can be written in the standard form,
\begin{widetext}
\begin{equation}
    \frac{\dd{\rho_S(x,x',t)}}{\dd{t}} = \frac{i\hbar}{2m} \left(\frac{\del^2}{\del x^2} - \frac{\del^2}{\del x'^2}\right)\rho_S(x,x') - \frac{\Lambda}{\hbar}(x-x')^2\rho_S(x,x').
\end{equation}
It is more convenient to work in the rotated coordinate system given by, $y=x-x'$ and $z=x+x'$.
The JZME thus becomes,
\begin{equation}
    \label{eqn:JZME}
    \frac{\dd{\rho_S(y,z,t)}}{\dd{t}} = \frac{2i\hbar}{m} \frac{\del^2}{\del y \del z} \rho_S(y,z,t) - \frac{\Lambda}{\hbar} y^2 \rho_S(y,z,t).
\end{equation}
The evolution of solutions to the JZME are characterised by the narrowing of the distribution of non-zero density matrix entries around $y=0$, i.e. the diagonal.
The key insight of CWT is that this single master equation can be written as two equations for the conditional and marginal wavefunctions, $\phi(x,q,t)$ and $a(x,t)$, respectively, where the total system-environment wavefunction, $\psi(x,q,t)$ is given by,
\begin{equation}
    \label{eqn:CWTfact}
    \psi(x,q,t)=\phi(x,q,t)a(x,t),
\end{equation}
where the generalised coordinate $q$ represents the degree(s) of freedom of the environment. The reduced density matrix is then given by,
    \begin{align}
    \begin{split}
        \rho_S(x,x',t) &= \text{tr}_E ~ \psi^*(x',q',t)\psi(x,q,t)\\
        &= \int \dd{q} ~ \psi^*(x',q,t) \psi(x,q,t).
    \end{split}
    \end{align}
Substituting in the CWT factorisation, we find,
\begin{equation}
    \rho_S(x,x',t) = a^*(x',t)a(x,t) \int \dd{q} ~ \phi^*(x',q,t)\phi(x,q,t) = \rho_M(x,x',t) K(x,x',t)
\end{equation}
where we've defined the \textit{marginal density matrix}, $\rho_M(x,x',t)=a^*(x',t)a(x,t)$ and the \textit{coherence integral} $K(x,x',t)=\int \dd{q} ~ \phi^*(x',q,t)\phi(x,q,t)$.
Substituting this back into the JZME (Eq. \ref{eqn:JZME}), we find the marginal density matrix and the coherence integral must satisfy,
\begin{equation}\label{eqn:factoredME}
    \frac{\dd{\rho_M}}{\dd{t}} = \frac{2i\hbar}{m}\left( \del_{yz}\rho_M + (\del_{y}\rho_M)(\del_{z} \ln K) + (\del_z\rho_M)(\del_{y} \ln K) + \rho_M\frac{\del^2_{yz}K}{K} \right) - \frac{\Lambda}{\hbar}y^2 \rho_M - \frac{\dd{}}{\dd{t}} \ln K.
\end{equation}
However, CWT also requires the marginal density matrix obey an equation of motion of its own \cite{vang2020}. This arises from the equation of motion for the marginal wavefunction, which itself arises from the Schr\"{o}dinger equation, the CWT factorisation, Eq. \eqref{eqn:CWTfact}, and the requirement,
\begin{equation}
    \int \dd{q} \, \phi^*(x,q,t) \phi(x,q,t) = 1.
\end{equation}
With an appropriate choice of gauge, the equation of motion for the marginal wavefunction can be written as,
\begin{equation}
\label{eqn:marginaleom}
    i\frac{\dd{a}(x,t)}{\dd{t}} = -\frac{\hbar}{2m} \frac{\del^2 a(x,t)}{\del x^2} + \frac{1}{\hbar}\eps(x,t)a(x,t)  
\end{equation}
where $\eps(x,t)$ is a real-valued function. This leads to an equation of motion for the marginal density matrix through the relation, $\rho_M(x,x',t) = a^*(x',t)a(x,t)$,
\begin{equation}
\label{eqn:marginalME}
    \frac{\dd{\rho_M}}{\dd{t}} = \frac{i\hbar}{2m} \left(\del^2_x - \del^2_{x'}\right) \rho_M - \frac{i}{\hbar}\left[\eps(x,t)-\eps(x',t)\right]\rho_M.
\end{equation}
Our aim then, is to find a form of $\eps(x,t)$ and $K(x,x',t)$ such that Eq. \eqref{eqn:marginalME} and Eq. \eqref{eqn:factoredME} are equivalent. One such form can be found by noting the fact that the original JZME has analytic Gaussian solutions, suggesting a Gaussian form for $K$ may satisfy this requirement. In particular, if we take,
\begin{equation}
    K(y,t) = \exp\left(-\frac{\gamma(t) y^2}{2}\right)
\end{equation}
where we're working in the $(y,z)$ basis for convenience, then it can be verified by substitution that Eq. \eqref{eqn:factoredME} and Eq. \eqref{eqn:marginalME} coincide as $y\rightarrow 0$ provided,
\begin{equation}
    \eps(x,t) = \frac{\hbar^2}{m} \gamma(t) \ln|a(x,t)|^2.
\end{equation}
For $y \neq 0$, the two equations coincide only if,
\begin{equation}
\label{eqn:lingamma}
    \dv{\gamma(t)}{t} = \frac{\Lambda}{\hbar} \implies \gamma(t) = \gamma(0) + \Lambda t/\hbar.
\end{equation}
However, since all observables depend only on the diagonal values of $\rho_M(x,x')$, that is, when $x=x'$ or equivalently $y=0$, the requirement that the evolution of the off-diagonal terms is not strictly necessary. In the Gaussian case for example Eq. \eqref{eqn:lingamma} does not hold for all time, but still shows good agreement with the JZME, as discussed in the main text.
A more general solution can be found by dropping the Gaussian assumption for $K(x,x',t)$. The expression for $\eps(x,t)$ in that case has the form,
\begin{equation}
\label{eqn:generaleps}
    \eps(x,t) = C(t) + \int_0^{x} \dd{x'} \gamma_2(x',t) \pdv{}{x'}  \ln[\gamma_2(x',t)r(x',t)]
\end{equation}
where $r(x,t)=|a(x,t)|$ and $C(t)$ is an arbitrary function of time which does not effect the dynamics, and where $\gamma_2(x,t)$ is defined as,
\begin{equation}
    \gamma_2(x,t) = \frac{\del^2}{\del y^2} \ln K(y,z,t)\Big|_{y=0}.
\end{equation}
Equation Eq. \eqref{eqn:generaleps} can be verified by substitution of
\begin{equation}
    \gamma(t)y^2/2 \rightarrow \sum_n\gamma_n(z,t)y^n/n!
\end{equation} 
in equation \ref{eqn:marginalME}.
This agrees with the Gaussian case exactly when $\pdv{\gamma_2}{x}=0$, which is to say that $\gamma_n=0 $ for $n>2$.
\end{widetext}
\bibliography{main.bib}
\end{document}